# Deep learning vessel segmentation and quantification of the foveal avascular zone using commercial and prototype OCT-A platforms

**Short Title:** Quantitative OCT-A with DNN

Morgan Heisler[1,¶] MASc, Forson Chan[2,¶] MD, Zaid Mammo[3,*] MD, Chandrakumar Balaratnasingam[4,5,6,7] MD, Pavle Prentasic[8] PhD, Gavin Docherty[3] MD, MyeongJin Ju[1] PhD, Sanjeeva Rajapakse[2] MD, Sieun Lee[1] PhD, Andrew Merkur[3] MD, Andrew Kirker[3] MD, David Albiani[3] MD, David Maberley[3] MD, K. Bailey Freund[6,7] MD, Mirza Faisal Beg[1] PhD, Sven Loncaric[8] PhD, Marinko V. Sarunic[1] PhD, and Eduardo V. Navajas[3] MD, PhD

[1] School of Engineering Science, Simon Fraser University, Burnaby, British Columbia, Canada
[2] Faculty of Medicine, University of British Columbia, Vancouver, British Columbia, Canada
[3] Department of Ophthalmology and Visual Sciences, University of British Columbia, Vancouver, British Columbia, Canada
[4] Centre for Ophthalmology and Visual Science, University of Western Australia, Perth, Western Australia, Australia
[5] Lions Eye Institute, University of Western Australia, Perth, Western Australia, Australia
[6] Vitreous Retina Macula Consultants of New York, New York, United States
[7] Manhattan Eye, Ear and Throat Hospital, LuEsther T. Mertz Retinal Research Center, New York, New York, United States
[8] University of Zagreb, Faculty of Electrical Engineering and Computing, Zagreb, Croatia

*Corresponding author: Dr. Zaid Mammo
Address: 2550 Willow St, Vancouver, BC V5Z 3N9
E-mail: znmammo@gmail.com
Phone: +61 432 147 437

¶These authors contributed equally to this work

**Commercial relationships:** K Bailey Freund Consultant to: Genentech, Optos, Optovue, Heidelberg Engineering, and Graybug Vision; Research support: Genentech/Roche.

# Abstract

**Objective:** Automatic quantification of perifoveal vessel densities in optical coherence tomography angiography (OCT-A) images face challenges such as variable intra- and inter-image signal to noise ratios, projection artefacts from outer vasculature layers, and motion artefacts. This study demonstrates the utility of deep neural networks for automatic quantification of foveal avascular zone (FAZ) parameters and perifoveal vessel density of OCT-A images in healthy and diabetic eyes.

**Methods:** OCT-A images of the foveal region were acquired using three OCT-A systems: a 1060nm Swept Source (SS)-OCT prototype, RTVue XR Avanti (Optovue Inc., Fremont, CA), and the ZEISS Angioplex (Carl Zeiss Meditec, Dublin, CA). Automated segmentation was then performed using a deep neural network. Four FAZ morphometric parameters (area, min/max diameter, and eccentricity) and perifoveal vessel density were used as outcome measures.

**Results:** The accuracy, sensitivity and specificity of the DNN vessel segmentations were comparable across all three device platforms. No significant difference between the means of the measurements from automated and manual segmentations were found for any of the outcome measures on any system. The intraclass correlation coefficient (ICC) was also good ($\geq 0.51$) for all measurements.

**Conclusion:** Automated deep learning vessel segmentation of OCT-A may be suitable for both commercial and research purposes for better quantification of the retinal circulation.

**Keywords**: optical coherence tomography, angiography, deep learning, diabetic retinopathy

# Introduction

Diabetic retinopathy (DR) is the most prevalent retinal vascular disease worldwide, affecting a third of people with diabetes(1). It ranks as the fifth most common cause of world-wide blindness and of global moderate to severe vision impairment(2). The pathophysiology of DR involves damage to the inner retinal microcirculation leading to the disruption of the blood retinal barrier and capillary bed closure (3,4). Studies have shown that retinal ischemia secondary to capillary non-perfusion occurs in the early stages of diabetic retinopathy and is associated with disease severity and progression (5). In addition, decreased macular capillary density and enlargement of the foveal avascular zone (6) showed correlation with visual acuity in DR patients (7,8). Therefore, quantitative assessment of vascular changes at the capillary level may result in early DR detection, better disease severity stratification and timely intervention before the onset of irreversible vision loss.

For the past 50 years, fluorescein angiography (FA) has been the gold standard modality for assessing retinal vascular diseases allowing accurate detection of large and medium sized vessels (9). However, it requires injection of a toxic dye and capillary visibility by FA is affected by poor contrast from background choroidal fluorescence and light scattering in the retina. This poor capillary visualization increases with retinal depth making FA less suitable for deep capillary plexus detection. (10). Optical coherence tomography angiography (OCT-A) is a depth-resolved, non-invasive, and dye-free imaging modality that uses flow-based contrast to visualize the multi-layered retinal circulation. By utilizing a custom built OCT-A prototype, our group has shown that OCT-A images provide histology-like anatomical information about human retinal capillary networks (11–13). Recently, several studies have shown favourable comparisons between OCT-A and FA imaging in normal subjects(14) and patients with retinal vascular disease including DR(15,16).

Early efforts aimed at OCT-A perifoveal vessel density quantification utilized manual vessel tracing methods(14), but the process can be labour intensive and subject to intra- and inter-observer variability(17). Fully automated techniques are being explored, but face challenges such as variable intra- and inter-image signal to noise ratios, projection artefacts from outer layer vasculature onto deeper layers, incorrect layer segmentation and motion artefacts(18–20). OCT-A signal intensity thresholding has been the foundation of most automated segmentation efforts thus far and progress has been made in applying additional filters for more accurate results(21,22). Improvements in automated segmentation and quantification of OCT-A images of the retinal vasculature may aid in its wider spread adoption and potential applications.

Our group has previously demonstrated a novel automated deep learning method to segment and quantify retinal images from our prototype OCT-A machine using deep neural networks (DNN)(19). In this study, we expand upon the application of the algorithm in several ways: We utilize larger fields of view that are more clinically useful (2x2mm and 3x3mm) and locations (centered on the FAZ) from the original 1x1mm (perifoveal) to retrain the DNN, validate the DNN for 2 commercial OCT-A systems, include OCT-A images from eyes with DR, and automatically quantify FAZ parameters. To our knowledge, this is the first paper to validate an automated retinal vessel segmentation of OCT-A images with manual segmentation across multiple platforms.

## Methods

The protocol for this study was approved by the human research ethics committees of Simon Fraser University, the University of British Columbia, and the North Shore Long Island Jewish Health System and conducted in compliance with the Declaration of Helsinki. Written informed consent was obtained from all subjects. Patient imaging using the commercial device was performed at Vitreous Retina Macula Consultants of New York from February 2015 to April 2016 and North

Shore Associates, North Vancouver, British Columbia from September 2016 to April 2017. Patient imaging using the prototype device was performed in the Advanced Retinal Imaging and Analysis Laboratory (ARIAL) at the Eye Care Centre in Vancouver, British Columbia from July 2014 to October 2016. Data analysis for this study was performed from February 2016, to August 2017. The subjects underwent a standard ophthalmic examination and their level of retinopathy was determined by the treating physician using the Early Treatment of Diabetic Retinopathy Study (ETDRS)(23) staging.

## *Inclusion criteria*

Subjects classified as diabetic were diagnosed with DR according to the ETDRS criteria by an experienced retina specialist. Subjects that comprised the control group showed no evidence of retinal or ocular pathology on examination. All subjects were screened for clear ocular media, ability to fixate, and competence to provide informed consent prior to imaging.

## *Optical coherence tomography instrumentation*

Three OCT-A systems were used in this study: one prototype and two commercially available machines. The clinical prototype OCT-A system used in this study was designed and built by the Biomedical Optics Research Group (BORG) at Simon Fraser University using a 1060 nm swept-source (Axsun Inc, Billerica, MA) with a 100 kHz A-scan rate. The axial resolution was ~6 μm in tissue and the estimated focal waist ($1/e^2$ Gaussian radius) was ~7.3 μm at the retina. The details of this acquisition system have previously been published(24). One commercial OCT-A system used in this report was the RTVue XR Avanti (Optovue Inc., Fremont, CA) which is an 840 nm spectral domain system with an A-scan rate of 70 kHz. The XR Avanti has a reported axial resolution of 5μm and a transverse resolution of 15μm(25). A second commercial OCT-A system used was the ZEISS Angioplex (Carl Zeiss Meditec, Dublin, CA) which also uses an 840 nm spectral domain source and has an A-scan rate of 68 kHz. The reported axial and transverse resolution is 5 μm and 15 μm respectively(26).

## *Imaging protocols*

Standard imaging procedures differed among the three OCT-A systems. For the prototype OCT-A system, three repeat acquisitions at each B-scan location were acquired over a 2x2mm area centered on the Foveal Avascular Zone (FAZ) sampled in a 300×300 (×3) grid, which required ~3.15s for image acquisition. The real-time speckle variance (sv) OCT processing and display of the OCT-A images was performed using our open source software(27,28). For the RTVue-XR Avanti system, images were scanned over 3x3mm regions centered on the FAZ with a scan pattern of 2 repeated B-scans at 304 raster positions, with each B scan consisting of 304 A-scans. Two volumetric scans were acquired in this fashion: one horizontally scanned and the other vertically for a total acquisition time of ~6s(25). For the Zeiss Angioplex system, the 3 x 3 mm scan protocol was used, which consisted of 245 A-scans per B-scan at 245 B-scan positions. Each B-scan was repeated 4 times for each position(26). A summary of these protocols is shown in Fig 1.

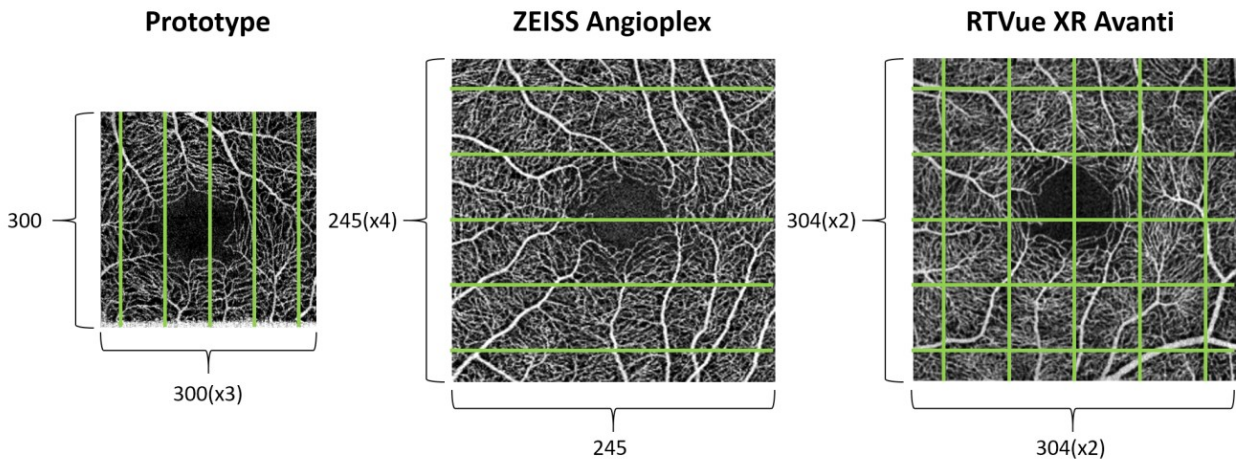

**Fig 1. Imaging protocols for the three different OCT-A systems.**
Green lines indicate the fast scanning direction. The number of locations sampled are labelled for each axis where the numbers in brackets (x#) indicate the number of repeated B-scans per location.

## Processing of OCT-A images

Post-processing of the image data acquired with the 1060nm prototype OCT-A system started with automated motion correction. Coarse axial motion artefact was corrected using cross-correlation between adjacent frames then translational sub-pixel registration(29) was then performed on each set of corresponding B-scans to further eliminate any inter-scan motion artefact. The inner limiting membrane (ILM) and posterior boundary of the outer plexiform layer (OPL) were then segmented automatically in 3D using an automated graph-cut algorithm(30). Automated segmentations were examined and corrected when necessary by a trained researcher using Amira (version 5.1; Visage Imaging, San Diego, CA, USA). The OCT-A scans were then summed in the axial direction between the segmented layers to produce a projected *en face* image. The *en face* images were then automatically notch filtered and contrast-adjusted using adaptive histogram equalization.

Both commercial images were processed with the systems' built-in image processing software. Briefly, for the Optovue system, the split-spectrum amplitude-decorrelation angiography (SSADA) method was used for extracting the OCT-A information. The algorithm split the spectrum into 11 sub-spectra and detected blood flow by calculating the signal amplitude-decorrelation between two consecutive B-scans of the same location. Both horizontally acquired and vertically acquired images were registered and averaged. For the Zeiss system, OCT microangiography (OMAG) was used as the OCT-A algorithm, which incorporates both the phase and intensity information of the B-scans. The inner limiting membrane (ILM) and posterior boundary of the outer plexiform layer (OPL) were also used as the segmentation boundaries for both commercial devices as well (26).

## Manual tracing methods

One masked, trained rater (FC) segmented all OCT-A images using a Samsung ATIV Smart PC Pro 700T tablet (FC) and GNU Image Manipulation Program (GIMP). The segmentations were reviewed and accepted by two other trained raters (MH and ZM).

## Algorithm training methods

The automated segmentation of the blood vessels in the OCT-A images was performed by classifying each pixel into vessel or non-vessel class using deep convolutional neural networks. A detailed description of the DNN architecture has been previously published(19). Briefly, original OCT-A *en face* images and the corresponding manual segmentations were used as inputs to train the deep neural network. An equal number of vessel and non-vessel pixels were extracted from each image to create a balanced training set. The trained network then segmented the test datasets by assigning a grayscale value, with higher values representing higher confidence of the pixel being a vessel. The prototype and commercial devices were trained separately due to inherent differences in the original images such as the image size. All datasets were trained on a mixed training set comprised of both healthy and diabetic images. The first half of the dataset was used to train the deep neural network, which was then used to segment the second half of the dataset. The process was repeated with the datasets reversed. Due to an icon in the bottom left corner of some data, a mask was applied over the area and it was disregarded in further analysis.

## Segmentation performance analysis

The automated deep learning segmentation was thresholded using Otsu's method to create a binary output(31). Segmentation performance was then evaluated by pixel-wise comparison against the manually segmented images. The accuracy, sensitivity, and specificity were calculated for each pixel in the dataset and presented as a mean average. For each dataset the number of true positive (TP), true negative (TN), false positive (FP), and false negative (FN) pixels were used to calculate the accuracy ((TP+TN)/(TP+FP+FN+TN)), sensitivity ((TP/(TP+FN)), and specificity (TN/(TN+FP)).

### *Clinical outcome measures*

Four FAZ morphometric parameters (area, maximum and minimum diameter, and eccentricity) as well as perifoveal vessel density were calculated from the automated segmentation results. The FAZ was found as the largest connected non-vessel area. The centroid for this area was then used to determine the maximum and minimum diameter. Eccentricity was calculated as $e = \sqrt{1 - \frac{b^2}{a^2}}$ where $b$ is the minimum radius and $a$ is the maximum radius of the ellipse made by the maximum and minimum diameter. Before capillary density measurements were calculated for the automated segmentations a gamma correction filter was applied to ensure vessel connectivity after binarization. Additionally, all erroneously segmented pixels within the FAZ area were set to a non-vessel classification. Perifoveal vessel density was then calculated as the proportion of measured area occupied by pixels which were classified by the algorithm as a vessel.

Paired t-tests and Intraclass Correlation Coefficients (ICC) were used to compare the means and agreement between segmentation methods, respectively, of the four FAZ morphometric parameters and perifoveal vessel density. A Student's t-test assuming heteroscedasticity was used to compare automatically segmented eyes with and without DR. Results for the prototype and commercial OCT-A systems were assessed separately.

## Results

### *Demographics*

A total of 81 eyes from 46 subjects were imaged as per the study protocol. Both eyes were imaged for all subjects imaged with the prototype and Optovue system. Three eyes were excluded from both the control and diabetic groups of the prototype system due to excessive motion artefacts (5 or more per image). Only one eye (chosen randomly) was imaged per subject with the ZEISS
Angioplex system, with the exception of 3 control subjects and 2 diabetic subjects. The
demographics for all study subjects are included in Table 1.

**Table 1. Subject Demographics and Clinical Characteristics**

|  | Prototype System | Optovue RTVue XR Avanti | ZEISS Angioplex |
|---|---|---|---|
| Subjects (n) | | | |
|   Control | 12 | 8 | 10 |
|   Diabetic | 5 | 4 | 8 |
| Eyes (n) | | | |
|   Control | 21 | 16 | 13 |
|   Diabetic | 7 | 8 | 10 |
| ETDRS Grade (n) | | | |
|   Mild NPDR | 1 | 1 | 3 |
|   Moderate NPDR | - | 2 | 3 |
|   Severe NPDR | 2 | - | 1 |
|   Early PDR | 4 | 5 | 1 |
| Age (mean ± std) | 37.3 ± 12.5 | 39.3 ± 13.3 | 52.7 ± 17.0 |
| Gender (M:F) | 10:7 | 6:6 | 8:9 |

## *Deep Neural Network algorithm performance*

An example of the automated segmentation output for both the prototype and commercial systems is shown in Fig 2 along with the corresponding original image and manual segmentation. For the images acquired with the Optovue system, the accuracy (healthy: 0.83, diabetic: 0.87), sensitivity (healthy: 0.82, diabetic: 0.86), and specificity (healthy: 0.84, diabetic: 0.88) of the deep learning algorithm were calculated. The Zeiss system's accuracy (healthy: 0.84, diabetic: 0.85), sensitivity (healthy: 0.84, diabetic: 0.87), and specificity (healthy: 0.83, diabetic: 0.81) were similar. The fields of view for the 1060nm prototype system was 2x2mm instead of 3x3mm, but the accuracy (healthy: 0.797, diabetic: 0.833), sensitivity (healthy: 0.806, diabetic: 0.733), and specificity (healthy: 0.790, diabetic: 0.881) were still similar. Although the algorithm performed quite well on most of the images, common OCT-A image quality issues such as low signal-to-noise ratio within the FAZ and motion artefacts, as seen in Fig 3, caused erroneous segmentations in some cases.

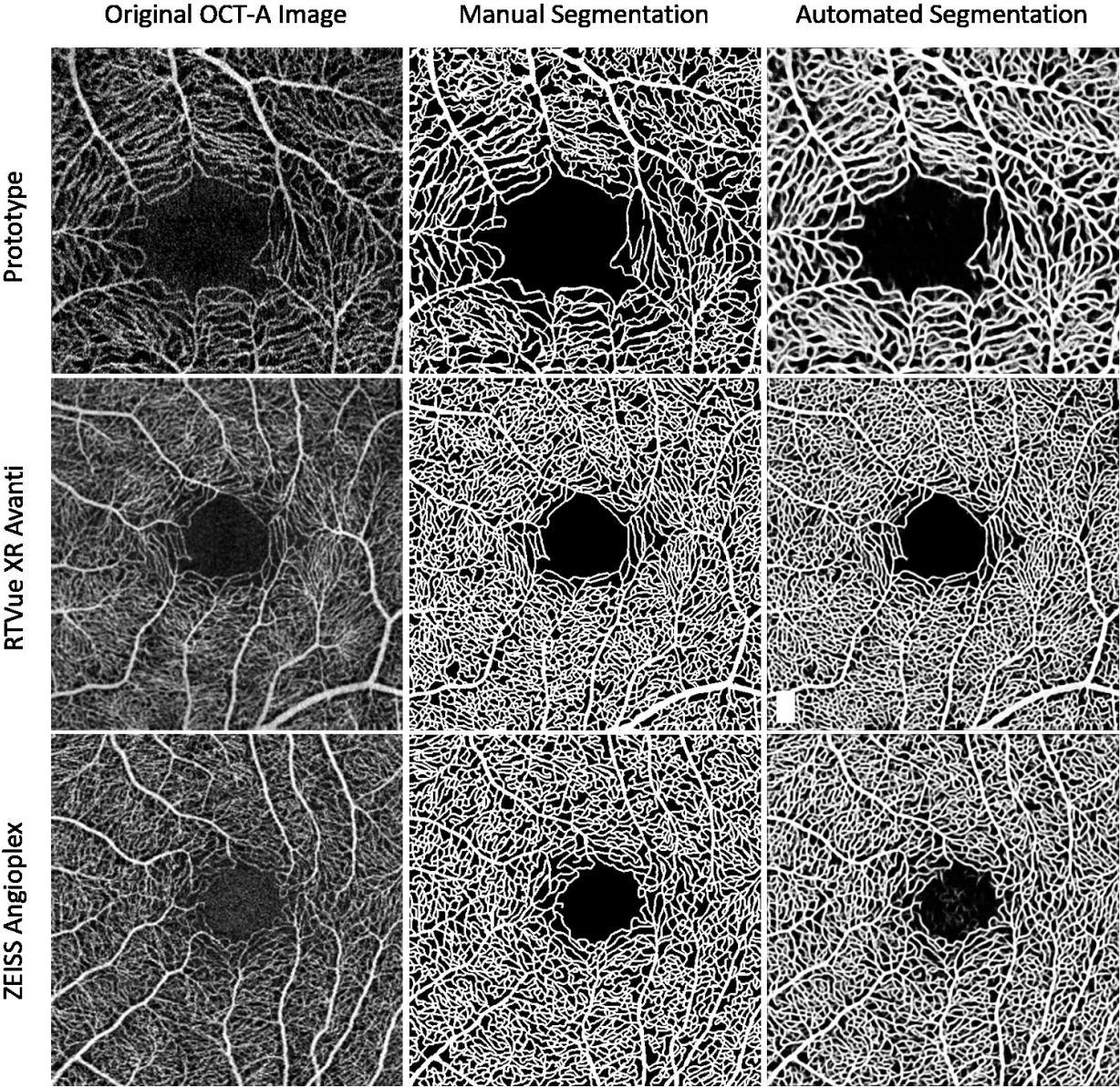

**Fig 2. Sample automated segmentations of OCT-A images from control eyes using a deep neural network.**
As some data within the Optovue RTVue XR Avanti dataset contained an icon in the lower left corner, a mask was applied and can be seen in the lower left corner of the automated segmentation results (white).

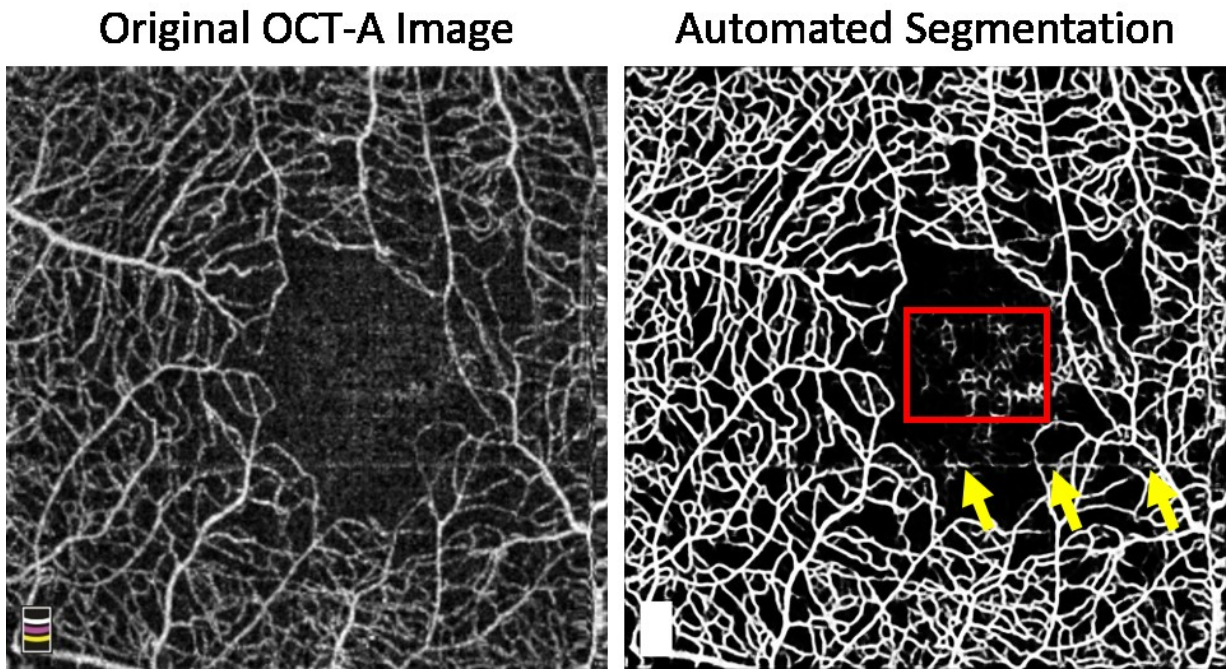

**Fig 3. Sample images showing the effects of low quality input data on the automated segmentation.**
Due to the low signal-to-noise ratio within the FAZ, some areas (for example the area within the red box) were erroneously segmented. Additionally, a horizontal motion artefact can be seen cutting through the FAZ which was also incorrectly segmented in areas (yellow arrows).

## *Clinical outcome measures*

Representative control images from all systems using both segmentation methods for the FAZ perimeter, minimum diameter, and maximum diameter are shown in Fig 4. Table 2 shows the results for the clinical outcome measures in both systems. No significant difference existed between the means of the clinical parameters derived from the manual and automated segmentations of images from the OCT-A systems. All statistical measures are reported in Table 2.

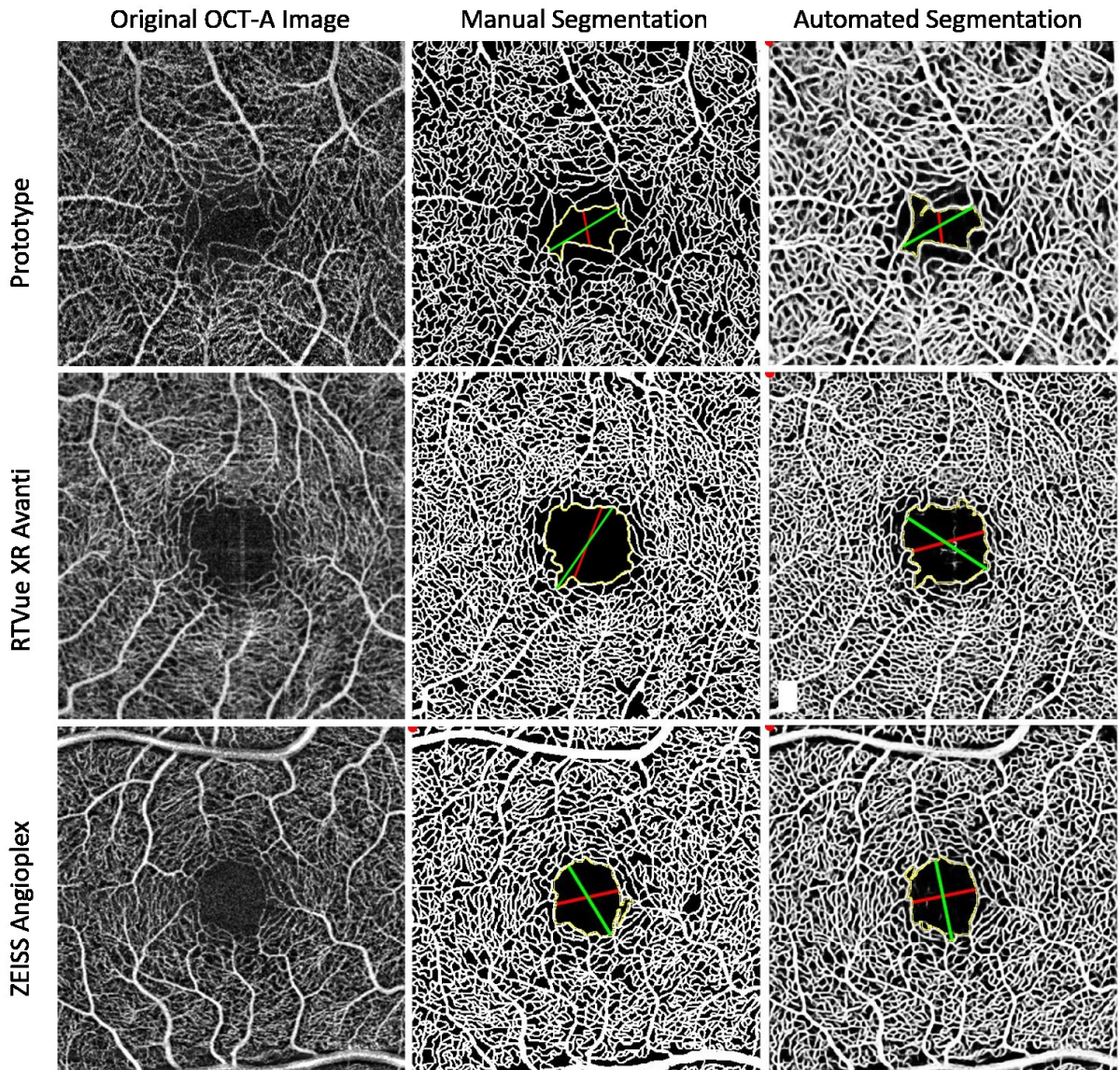

**Fig 4. Example segmentations and FAZ parameterization of control eyes for all three systems in the study.** FAZ parameters such as FAZ perimeter (yellow), maximum diameter (green) and minimum diameter (red) are shown for example healthy from 3 OCT-A systems using both manually and automated segmentations. As some data within the Optovue RTVue XR Avanti dataset contained an icon in the lower left corner, a mask was applied and can be seen in the lower left corner of the automated segmentation results (white).

**Table 2. Clinical Outcomes Measures for Comparing Automated and Manual Vessel Segmentation**

| | | FAZ Area (mm$^2$) | Minimum FAZ Diameter (mm) | Maximum FAZ Diameter (mm) | FAZ Eccentricity | Perifoveal vessel Density (%) |
|---|---|---|---|---|---|---|
| **Zeiss OCT-A** | **Healthy (n = 13)** | | | | | |
| | Manual | 0.278 ± 0.130 | 0.478 ± 0.116 | 0.736 ± 0.214 | 0.750 ± 0.052 | 0.493 ± 0.026 |
| | Automated | 0.332 ± 0.154 | 0.500 ± 0.102 | 0.748 ± 0.172 | 0.730 ± 0.064 | 0.513 ± 0.034 |
| | T-test | p = 0.35 | p = 0.60 | p = 0.90 | p = 0.41 | p = 0.11 |
| | ICC | 0.96 | 0.97 | 0.96 | 0.51 | 0.94 |
| | **Diabetic (n = 10)** | | | | | |
| | Manual | 0.552 ± 0.416 | 0.539 ± 0.217 | 1.035 ± 0.436 | 0.837 ± 0.056 | 0.378 ± 0.069 |
| | Automated | 0.736 ± 0.480 | 0.514 ± 0.166 | 1.164 ± 0.354 | 0.886 ± 0.047 | 0.418 ± 0.058 |
| | T-test | p = 0.37 | p = 0.78 | p = 0.48 | p = 0.06 | p = 0.18 |
| | ICC | 0.93 | 0.96 | 0.90 | 0.74 | 0.88 |
| **Optovue OCT-A** | **Healthy (n = 16)** | | | | | |
| | Manual | 0.280 ± 0.098 | 0.514 ± 0.114 | 0.704 ± 0.123 | 0.677 ± 0.087 | 0.519 ± 0.042 |
| | Automated | 0.261 ± 0.080 | 0.472 ± 0.096 | 0.689 ± 0.123 | 0.717 ± 0.089 | 0.525 ± 0.014 |
| | T-test | p = 0.69 | p = 0.27 | p = 0.86 | p = 0.16 | p = 0.42 |
| | ICC | 0.94 | 0.99 | 0.89 | 0.85 | 0.54 |
| | **Diabetic (n = 8)** | | | | | |
| | Manual | 0.594 ± 0.347 | 0.634 ± 0.226 | 1.114 ± 0.363 | 0.792 ± 0.097 | 0.385 ± 0.053 |
| | Automated | 0.492 ± 0.249 | 0.537 ± 0.159 | 0.950 ± 0.273 | 0.810 ± 0.089 | 0.418 ± 0.048 |
| | T-test | p = 0.51 | p = 0.34 | p = 0.32 | p = 0.70 | p = 0.25 |
| | ICC | 0.97 | 0.96 | 0.92 | 0.90 | 0.98 |
| **Prototype OCT-A** | **Healthy (n = 21)** | | | | | |
| | Manual | 0.300 ± 0.146 | 0.485 ± 0.174 | 0.707 ± 0.229 | 0.721 ± 0.083 | 0.381 ± 0.043 |
| | Automated | 0.279 ± 0.113 | 0.470 ± 0.130 | 0.707 ± 0.159 | 0.740 ± 0.092 | 0.388 ± 0.015 |
| | T-test | p = 0.11 | p = 0.58 | p = 0.97 | p = 0.34 | p = 0.54 |
| | ICC | 0.96 | 0.89 | 0.87 | 0.85 | 0.68 |
| | **Diabetic (n = 7)** | | | | | |
| | Manual | 0.415±0.164 | 0.516 ± 0.161 | 0.928 ± 0.164 | 0.819 ± 0.083 | 0.311±0.047 |
| | Automated | 0.366±0.104 | 0.446 ± 0.088 | 0.895 ± 0.172 | 0.852 ± 0.052 | 0.281±0.015 |
| | T-test | p=0.23 | p=0.13 | p=0.52 | p=0.30 | p=0.09 |
| | ICC | 0.89 | 0.83 | 0.86 | 0.62 | 0.63 |

The mean (± STD) of the clinical outcome parameters (FAZ area, minimum diameter, maximum diameter, eccentricity and perifoveal vessel density) are shown for both healthy and diabetic eyes and both OCT-A systems. The p-value from the paired t-test and the Intraclass Correlation Coefficient (ICC) value is also shown to compare manual and automated methods.

For all three OCT-A systems, eyes with DR had significantly lower perifoveal vessel density, greater maximum diameter, and greater FAZ eccentricity compared to the healthy eyes (p<0.05). Eyes with DR and normal eyes did not have any statistically significant differences in the FAZ area or minimum diameter in any system. Representative diabetic images from all systems using both segmentation methods for the FAZ perimeter, minimum diameter, and maximum diameter are shown in Fig 5.

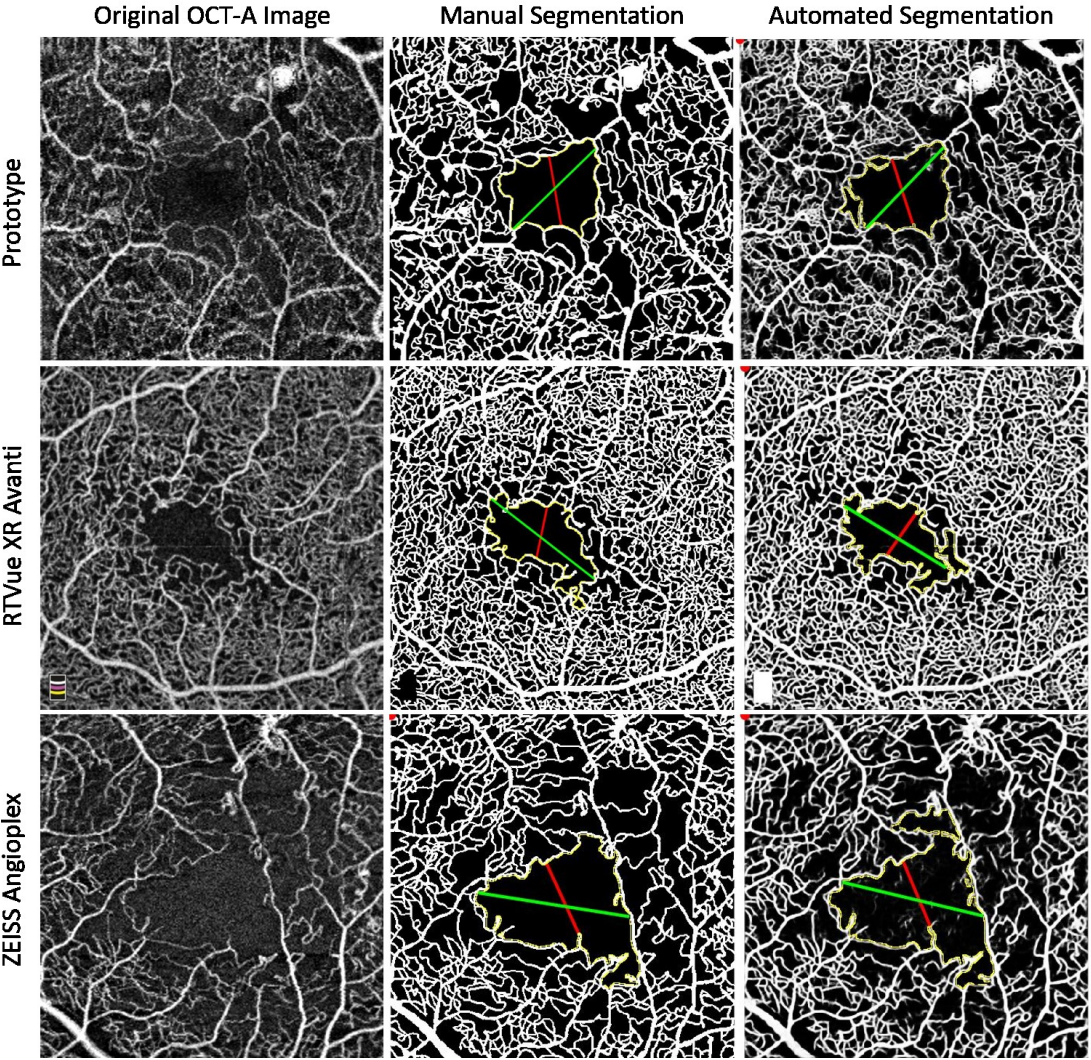

**Fig 5. Diabetic eyes have larger maximum diameter and eccentricity.** FAZ parameters such as FAZ perimeter (yellow), maximum diameter (green) and minimum diameter (red) are shown for example diabetic data from 3 OCT-A systems using both manually and automated segmentations. As the data within the Optovue RTVue XR Avanti dataset contained an icon in the lower left corner, a mask was applied and can be seen in the lower left corner of the manual segmentation (black) and automated segmentation results (white).

# Discussion

This study demonstrates the ability of a machine-learning based automated segmentation algorithm to segment the vessels of both healthy and diabetic eyes imaged with prototype and commercial OCT-A devices. The major findings of the study are: 1) Pixel-wise, the accuracy of the automated segmentation was comparable to that of a manual rater in all three OCT-A platforms, 2) Deep learning based segmentation can reliably quantify the perifoveal vessel density compared to manual segmentation across multiple OCT-A platforms and 3) Deep learning based segmentation has the capacity to reliably quantify the FAZ area, eccentricity, maximum and minimum diameter compared to manual raters across OCT-A platforms.

The fovea centralis is the anatomical area responsible for the highest visual acuity. With the exception of the foveola, the metabolic demands of the fovea centralis are met by a unique arrangement of superficial retinal capillaries(32). These end-arterial capillaries lack anastomoses which make this retinal eccentricity especially vulnerable to ischemic insult by retinal vascular diseases, including DR(33). Perifoveal vessel ischemia and FAZ enlargement are well-documented observations of macular ischemia in DR and are correlated to disease severity and progression(6–8). While FA remains the current clinical standard for evaluating macular ischemia, its invasive nature and potential adverse events makes it challenging to incorporate in regular screening and frequent follow-up of patients with DR. OCT-A is an alternative non-invasive, label-free imaging modality that has been favorably compared to histological representation and FA in the visualization of perifoveal circulation in healthy subjects and patients with DR. Accurate methods of quantification and analysis of OCT-A images are of great research and clinical interest. Optovue's built-in automated vessel segmentation extrapolates skeletonized outputs to show vessel density maps (34,35). Potential limitations to this method include underestimating vessel density in areas with thicker vessels and decreased sensitivity to capillary dropout(36). A more recently published approach by Schottenhamml et al.(36) takes advantage of a 'vesselness' filter to exploit the interconnective nature of the retinal vasculature and generate more detailed vessel segmentations; however, inaccuracies seem to result with vessel shape at vertices. The deep learning based method used in this report takes advantage of the 2-dimensional spatial structure of training images and was able to accurately mimic manual segmentations. Pixel-wise, the accuracy of the automated segmentation outputs compared to manual ranged from 76-91%, which is comparable to a reported inter-rater manual segmentation of ~83% agreement(19).

Perifoveal vessel non-perfusion is defined as the pathological disruption of perifoveal circulation with subsequent enlargement of inter-vessel distances between perifoveal vessel networks(7). It has been suggested that early perifoveal vessel non-perfusion could be present and represent an important biomarker in the absence of obvious DR on clinical examination(37) OCT-A has been shown to be able to delineate the perifoveal vessel networks precisely and consistently in patients with DR(38). This serves as a motivation to develop an automated tool that can accurately and reliably quantify the perifoveal vessel density. We found no statistically significant difference in the means of the perifoveal vessel densities when comparing measurements derived from the automated and manual segmentations. In comparing diabetic and healthy eyes, the automated outputs from both systems found a significantly lower ($p<0.01$) perifoveal vessel density in the DR eyes. This suggests that perifoveal vessel density calculated using deep learning based segmentation might be a clinically useful tool in the evaluation and follow-up of ischemic changes in DR.

The FAZ approximately delineates the location of the foveola within the fovea centralis. The absence of retinal vasculature is believed to help optimize the image on central cones(39). Increased FAZ area has been well-correlated with decreased visual acuity(8) and the severity of capillary nonperfusion(5,7) in patients with DR. Although there is high inter-individual variability in FAZ metrics for healthy eyes, longitudinal progression of the FAZ morphology may be a useful biomarker for DR(21). Improved visualization of the FAZ, enabled by OCT-A has allowed researchers to further study this area as it relates to DR(40), retinal vein occlusion(41), and aging(34). To assess the clinical utility of the automated segmentations in calculating FAZ morphometric parameters, the FAZ area, minimum and maximum diameter, and eccentricity were calculated using both the manual and automated segmentations. No significant difference existed between the means of any morphometric parameters derived from the manual and automated segmentations. For all three systems, the diabetic eyes were found to have a greater FAZ maximum diameter, and greater eccentricity compared to the healthy eyes while no significant difference in FAZ area or minimum diameter was noted. This lack of correlation is likely due to the high inter-individual variability. A non-circular FAZ shape has been shown to be a sensitive marker for early DR (42,43) and may be a more reliable biomarker as indicated by the greater eccentricity in DR eyes.

This study demonstrates the ability of a deep learning based automated segmentation algorithm to reliably segment the perifoveal microvasculature and provide clinically useful FAZ morphological measures. A limitation of this study is a restricted sample size for the diabetic groups. Additionally, as the performance of a deep learning based approach is limited by the quality of the training data, our automated segmentation performance is limited by image quality and manual vessel segmentations. Although the manual segmentations were reviewed by two trained raters in an attempt to mitigate manual segmentation error, the segmentations were reviewed on a holistic level whereas the machine learns on a pixel-by-pixel basis. Another limitation is the need for a database of segmented images for each field of view and each machine. Most commercial machines allow the imaging of various different fields of view (for example 3x3, 6x6, and 8x8mm), of which we only chose to use 2x2mm and 3x3mm to analyze in this study. Experimentally, performance variability in the vessel thickness occurred when the training set and the segmented images were from different fields of view, therefore a training set is needed for each field of view.

Systematic screening of people with diabetes has been shown to be a cost-effective approach for identifying potential vision loss(44). OCT-A is a promising new technology that has the potential to help guide earlier management decisions and prognosis. Deep learning automated segmentation of OCT-A may be suitable for both commercial and research purposes for better quantification of the retinal circulation in healthy subjects and in subjects with retinal vascular disease.

# Acknowledgements

The authors would like to thank the following agencies for funding: Michael Smith Foundation for Health Research, Natural Sciences and Engineering Research Council of Canada, Canadian Institutes of Health Research and National Health, and Brain Canada.